\documentclass[11pt]{article}
\usepackage{amsmath,epsfig,graphicx,amssymb}
\usepackage[margin=0.7in]{geometry}
\newcommand{\beq}{\begin{equation}}
\newcommand{\eeq}{\end{equation}}
\newcommand{\ben}{\begin{eqnarray}}
\newcommand{\een}{\end{eqnarray}}
\newcommand{\bes}{\begin{subequations}}
\newcommand{\ees}{\end{subequations}}
\newcommand{\bFig}{\begin{figure}}
\newcommand{\eFig}{\end{figure}}
\date{}
\begin{document}
\title{Quantum Mechanics and Quantum Information Science:\\ The Nature of $\Psi$}
\author{Partha Ghose\footnote{partha.ghose@gmail.com} \\
Centre for Astroparticle Physics and Space Science (CAPSS),\\Bose Institute, \\ Block EN, Sector V, Salt Lake, Kolkata 700 091, India.}
\maketitle

\centerline{\em Based on the Inaugural Talk given at IPQI held at Institute of Physics, Bubaneswar, February 17-28, 2014}
\begin{abstract}
An overview is given of the nature of the quantum mechanical wave function.
\end{abstract}
\section{The Paradoxes in Quantum Mechanics}
Right from the initial days of quantum mechanics a debate has been raging as to whether the wave function $\psi$ is ontic (state of reality) or epistemic (state of knowledge). This is closely related to the mystery of quantum measurements. As a result, quantum mechanics has been riddled with paradoxes and conundrums.

{\flushleft{\em Wave-particle Duality}}
\vskip 0.1in
When single particle states of electrons or photons are made to fall on a double-slit, one finds a pattern on a distant plate which is similar to an interference fringe with classical light when both slits are open but a single-slit diffraction pattern when only one slit is open. The first puzzle is that one finds an interference pattern with `particles', and the second puzzle is that the interference pattern disappears when one tries to figure out which slit the particle went through. This is the well known paradox of wave-particle duality. 
\vskip 0.1in
{\flushleft{\em The Schr\"{o}dinger Cat}}
\vskip 0.1in
In 1935 Schr\"{o}dinger wrote a paper \cite{schr} in which he introduced the famous cat paradox. The idea was to show that the uncertainties of the microscopic world (the time of decay of a radioactive substance) can get transferred to macroscopic objects like a cat, resulting in superpositions of incompatible states like `dead' and `alive' which are never seen.

\vskip 0.1in
{\flushleft{\em The EPR Paradox}}
\vskip 0.1in
In the same paper Schr\"{o}dinger dwelt extensively with the idea of the nonseparability of two quantum systems that are spatially separated and non-interacting, introduced by Einstein, Podolski and Rosen \cite{epr}. He coined the term `entanglement' and argued that it is the key feature of quantum mechanics that distinguishes it from classical physics. Entanglement results in individual entities losing their separate identities in a holistic entity, the entangled state. Measurement results in the `conditional disjunction' of this entity (i.e. dependent on the choice of basis) in an obscure process outside the ambit of quantum theory \cite{vN}. The essence of Einstein's argument \cite{ein} is much simpler than the arguments that appear in the famous EPR paper which was actually written by Podolski and in which, according to Einstein, the main point got obscured by pedantry. Einstein's argument can be put simply as follows. Let A and B be two qubits whose entangled wave function can be expressed in two incompatible bases as
\ben
\Psi(AB) &=& c_1 \psi_1(A) \psi_2(B) + c_2 \psi_2(A) \psi_1(B)\\
&=& c^\prime_1 \phi_1(A) \phi_2(B) + c^\prime_2 \phi_2(A) \phi_1(B)
\een
with $|c_1|^2 + |c_2|^2 = |c^\prime_1|^2 + |c^\prime_2|^2 =1$. If one is free to choose the basis in which a measurement will be made on A, then it is clear that one would obtain different wave functions $\psi(B)$ or $\phi(B)$ for B depending on the choice. Now, if one assumes (i) that the ontic state of B is not in any way influenced by what is measured on A (locality or separability assumption), and (ii) that every ontic state of a system is associated with a unique wave function (one-to-one correspondence), then there is a problem. One way to get around this problem is to give up condition (ii), i.e. to allow more than one wave function to be associated with a given ontic state of a system. Such an interpretation is called epistemic, and it implies that quantum mechanics is an incomplete theory. This is the interpretation that Einstein favoured. The alternative is to give up the assumption (i), i.e. locality {\em or} the separability of A and B. The first option results in nonlocality or `spooky' action-at-a-distance, but not necessarily the second one, as we will see later. Einstein found both these options unacceptable.

Already in the 1927 Solvay Conference Einstein had given a simple argument to show that even in the case of a single particle, quantum mechanics was incompatible with local realism \cite{bac}, which was a precursor to the Bell theorem \cite{bell1} for bipartite systems. Consider a plane single particle wave function incident on a screen with a small hole. After passing through the hole, the wavefunction spreads out on the other side of it in the form of a spherical wave, and is finally detected by a large hemispherical detector. The spherical wave function propagating towards the detector obviously shows no privileged direction. Einstein observed:
\begin{quote}
If $|\psi|^2$ were simply regarded as the probability that at a certain point a given particle is found at a given time, it could happen that {\em the same} elementary
 process produces an action in {\em two or several} places on the screen. But the interpretation, according to which the $|\psi|^2$ expresses the probability that
 {\em this} particle is found at a given point, assumes an entirely peculiar mechanism of action at a distance, which prevents the wave continuously distributed in 
space from producing an action in {\em two} places on the screen.
\end{quote}
Einstein later remarked that this `entirely peculiar mechanism of action at a distance' was in contradiction with the postulate of relativity. 

\section{Interpretations of Quantum Mechanics}
\vskip 0.1in
{\flushleft{\em Bohr and von Neumann}}
\vskip 0.1in
As a result of the above paradoxes, quantum mechanics has been interpreted in many ways over the years. The original interpretation was the one given by Niels Bohr who insisted that all measurements must result in unambiguous results, and hence one must use {\em classical} measuring apparatuses free from quantum uncertainties. The system S and the apparatus A get into an `unanalyzable whole' from which only statistically valid results can be inferred by a process that is shrouded in mystery. In this interpretation the classical world is presumed to exist independently of the quantum world, and is even required for its interpretation. If the quantum world is taken to underlie the classical world and be independent of it, this approach appears unsatisfactory, and von Neumann \cite{vN} tried to rectify this by treating the measuring apparatus also quantum mechanically. This in turn resulted in an entangled state $\rho(S,A)$, and he had to introduce projection operators to reduce this pure state to a mixed state $\hat{\rho}(S,A) = \sum_i \Pi_i \rho(S,A)\Pi_i$. The elements of this diagonal matrix represent the probabilities of the various measurement outcomes. This reduction of the state is not unitary, and von Neumann called this non-unitary, and hence non-quantum mechanical process, `process 1' which is as mysterious as the Bohr process. He called the unitary Schr\"{o}dinger evolution `process 2'. It is `process 1' and its nonlocality that most interpretations have tried to avoid or eliminate.

\newpage
{\flushleft{\em The Many Worlds Interpretation}}
\vskip 0.1in
Hugh Everett proposed an interpretation \cite{ev} that many have claimed to be the only one consistent with the mathematical formalism of quantum mechanics, namely the `relative state' interpretation which later on came to be known as the `many worlds' interpretation \cite{dewitt}. His intention was to eliminate `process 1' and have pure Schr\"{o}dinger evolution, and he achieved that by introducing a process of `splitting' of the wave function into branches all of which exist simultaneously. In other words, all possible outcomes of a measurement actually exist simultaneously, and hence no splitting actually takes place. This means that the cat is both alive and dead but they are in different branches of the universe, both of which are equally real but do not interact with each other. A cosmological version of this is the one proposed by Aguirre and Tegmark \cite{teg}. Some have argued that the splitting of the universe into multiple orthogonal branches is essentially a metaphysical hypothesis, and do not satisfy the principle of Occam's razor.  

\vskip 0.1in
{\flushleft{\em Hidden Variable Theories}}
\vskip 0.1in
As a response to the EPR argument that quantum mechanics is incomplete, several hidden variable theories were proposed to complete quantum mechanics. The idea was to introduce hidden variables underlying the observable world of quantum phenomena in a way so as to restore realism, and often also determinism, at the fundamental level and yet recover the quantum mechanical predictions by averaging over these hidden variables. However, most of these theories were subsequently ruled out by certain no-go theorems, as we will see later, except the one proposed by Bohm (and earlier by de Broglie) \cite{dbb}. In the de Broglie-Bohm theory, the trajectories of particles are introduced as ontic hidden variables, and they are piloted by the wave function that obeys the Schr\"{o}dinger equation. The theory is completely equivalent to quantum mechanics in its observable predictions, but it restores determinism and realism at the fundamental ontic level by giving up locality. In fact, Bohm often emphasized the inherent nonlocality of this interpretation to be the only real significance of quantum physics vis-a-vis classical physics.

\vskip 0.1in
{\flushleft{\em Other Interpretations}}
\vskip 0.1in
Some other interpretations have also been propsed such as the consistent histories interpretation \cite{griffiths}, the transactional interpretation \cite{cramer}, the modal interpretation \cite{modal}, relational quantum mechanics \cite{rov} and quantum Bayesianism \cite{fuchs}. Unfortunately I have been given a limited time in which it is impossible to do justice to them all.

\section{No-Go Theorems}

There are several no-go theorems that rule out most hidden variable theories. The first one of these was formulated by von Neumann. 
\vskip 0.1in
{\flushleft{\em The von Neumann Theorem}}
\vskip 0.1in
What von Neumann tried to prove is that dispersion free states, i.e. hidden variables, are impossible. He did that by first noting that any real linear combination of any two Hermitian operators represents an observable, and then making use of the assumption that the same linear combination of expectation values is the expectation value of the combination. Although this assumption is true of quantum mechanical states, von Neumann assumed it to be true of all dispersion free states as well, which turned out not to be the case, as was first shown by Bohm and later by Bell \cite{bell2}. Bohm's argument rested on the fact that von Neumann's assumption was applicable to a very narrow class of hidden variables but not to hidden variables which were contextual in the sense that the observables in the theory are not properties belonging to the observed system alone but also to the measuring apparatus. 

The limitations of other associated theorems like that of Jauch and Piron and of Gleason will be found in Bell's paper \cite{bell2}. 
\newpage
{\flushleft{\em Bell's Theorem}}
\vskip 0.1in
In 1964 Bell proved his famous theorem \cite{bell1} showing that local realism was incompatible with quantum mechanics. The proof uses bipartite entangled systems and is too well known to be repeated here. The final experimental verifications of the violations of Bell's theorem dealt a death blow to local hidden variable theories.

{\flushleft{\em The Kochen-Specker Theorem}}
\vskip 0.1in

Instead of considering locality as a criterion as Bell did, Kochen and Specker considered non-contextual hidden variables, i.e. variables that depend exclusively on the quantum system being measured and not on the measuring device \cite{ks}. This is in conformity with the classical notion that measurements only reveal values of pre-existing properties of a system, and that the value of a property of a system does not depend on prior measurements of other compatible properties. The Kochen-Specker theorem is based on the two axioms of (i) value definiteness (i.e. all observables have definite values at all times) and (ii) noncontextuality (i.e. the value of an observable is independent of its measurement context). Kochen and Specker showed that theories with such hidden variables are incompatible with quantum mechanics when the dimension of the Hilbert space is three or more. This theorem together with the previous ones ruled out all hidden variable theories except those involving contextual and nonlocal hidden variables. The prime example of such a theory is the de Broglie-Bohm theory. 

\section{The Quantum Information Age}

The age of quantum information processing had its birth in 1982 when Feynman showed that a classical Turing machine would experience an exponential slowdown when simulating quantum proceses but his hypothetical universal quantum simulator would not \cite{feynman}. Then in 1985 David Deutsch defined a universal quantum computer \cite{deut}, and in 1996 Seth Lloyd showed that a quantum computer can be programmed to simulate any local quantum system efficiently \cite{seth}. In quantum computing the analog of the classical unit of information, the bit, is a qubit which is a two-level quantum system, like the two states of polarization of a single photon which can be in a superposition of two states,
\beq
|\psi\rangle = \alpha |0\rangle + \beta |1\rangle
\eeq
with $|\alpha|^2 + |\beta|^2 = 1$. The breakthroughs in quantum algorithms came between 1992 and 1996 with Deutsch and Jozsa \cite{dz} proposing a quantum algorithm that could run exponentially faster than any deterministic classical algorithm. (This was improved by Cleve, Ekert, Macchiavello and Mosca \cite{cleve} in 1998.) Then came the Shor algorithm for integer factorization \cite{shor} which runs on polynomial time and is in the complexity class BQP, and the Grover search algorithm for an unsorted database with $N$ entries in $O(N^{1/2})$ time using $O(log N)$ storage space \cite{grover}.

Of fundamental importance in quantum computing is the quantum gate (or quantum logic gate) which is a basic quantum circuit operating on a small number of qubits. All quantum circuits can be built from these building blocks, like classical logic gates for conventional digital circuits. Quantum logic gates are reversible. Classical computing can be performed using only reversible gates. For example, the reversible Toffoli gate can implement all Boolean functions. With the help of its quantum equivalent, the quantum Toffoli gate, one can show that quantum circuits can perform all operations performed by classical circuits. The quantum gates, represented by unitary matrices, mostly operate on spaces of one or two qubits. The commonly used gates are the Hadamard gate, the Pauli-X, Y and Z gates represented by the Pauli matrices $\sigma_x,\sigma_y, \sigma_z$, phase shift gates, swap gates, controlled gates like CNOT and the Toffoli gate.

Since this conference is specially meant for quantum information processing methods, many special techniques that have been developed, such as quantum Fourier transforms, adiabatic computing, error correction codes, dense coding, etc will be discussed here in the next few days. Let me move on to mention certain fundamental theorems that have been proved by leveraging basic properties of the wave function. 

\vskip 0.1in
{\flushleft{\em No-Cloning Theorem}}
\vskip 0.1in
This theorem states that {\em quantum mechanics forbids the creation of identical copies of an arbitrary unknown quantum state} \cite{wzd}. To appreciate the real significance of this theorem, let us consider a classical analogue of this theorem. Given the result of only one flip of a possibly biased coin, one cannot simulate a second, independent toss of the same coin. This follows from the linearity of classical probability. The proof of the quantum no-cloning theorem has the same structure. So, in what sense is no-cloning a uniquely quantum result? The answer lies in restricting quantum states to pure states that are not convex combinations of other states. Such states are not pairwise orthogonal as classical pure states are. The {\em no-broadcast theorem} generalized the no-cloning theorem to mixed states \cite{nobroad}.
\vskip 0.1in
{\flushleft{\em No-Deleting Theorem}}
\vskip 0.1in
{\em Given two copies of an arbitrary quantum state, it is impossible to delete one of the copies} \cite{pati}. This is a consequence of the linearity of quantum mechanics. It is a time-reversed dual of the no-cloning theorem. Like the no-cloning theorem it has profound implications in quantum computing, quantum information theory and quantum mechanics in general. For instance,  violation of this theorem would make faster-than-light signals possible. 

Together with the no-cloning theorem, it implies the conservation of quantum information (which prevents the holographic principle for black holes) and the interpretation of quantum mechanics in terms of category theory \cite{cat}. 

Quantum information science comprises many areas of research such as quantum computers, quantum complexity theory, quantum cryptography, quantum teleportation, quantum error correction, quantum communication complexity, quantum dense coding, etc. Most of these issues will be discussed at this meeting in the next few days. I would therefore like to pass on to a very fundamental issue concerning the nature of $\psi$ raised by quantum information theory which regards information as the most fundamental organizing entity in nature. 

\section{Is $\psi$ Epistemic?}

As we have seen, the interpretation of a wave function in quantum mechanics has been of considerable debate since its inception.  Some have argued for an ontic (state of reality) interpretation while others, including Einstein, have preferred an epistemic (state of knowledge) interpretation. Quantum information theory favours the epistemic interpretation \cite{qi}. An advantage of such an interpretation is that a sudden change or collapse of the wave function can be interpreted as a Bayesian updating on receiving new information, thus avoiding the measurement problem and nonlocality. Fuchs, Mermin and Schack \cite{fuchs} have advocated this quantum Bayesian (Qbist) approach. The most important question, therefore, is: which of these alternatives is correct? 

Recently Pusey, Barrett and Rudolph \cite{pbr} have proved a no-go theorem with a couple of reasonable assumptions (preparation independence) to rule out an epistemic interpretation. Collbeck and Renner then showed that if measurement settings can be chosen freely, a system's wave function is in one-to-one correspondence with its elements of reality, thus ruling out any subjective interpretation \cite{cr}. On the other hand, Lewis {\em et al} \cite{pg} have shown that if one drops the preparation independence assumption and also slightly weakens the  definition of an epistemic state, an epistemic interpretation is possible. The situation is far from clear and has been recently reviewed by Leifer \cite{leifer}.

\section{How Quantum is Entanglement?}
An unexpected recent discovery is that classical optics displays some quantum-like features such as entanglement, as originally predicted by Spreeuw \cite{sp} and independently by Ghose and Samal \cite{g}. This emerging field has been reviewed with emphasis on the Hilbert space structure of classical polarization optics \cite{ghose}. What is the precise origin of such quantumness in patently classical phenomena?

It turns out that the answer lies in the classic works of Koopman \cite{K}, von Neumann \cite{vN2} and later Sudarshan \cite{sud} who developed a complete phase space theory of classical mechanics based on Hilbert spaces with commuting hermitian operators as observables. The dynamics is Liouvillian. In addition to position $q$ and momentum $p$, there are additional canonically conjugate but unobservable operators $\hat{\lambda}_q,\hat{\lambda}_p$ in this formalism with the commutation properties $[\hat{q}, \hat{\lambda}_q]=i=[\hat{p}, \hat{\lambda}_p]$ (no Planck constant!), all other pairs commuting among themselves. 

Since the classical KvNS wave functions are complex, the phase of the wave function must be unobservable in classical mechanics. This is inherent in the KvN formalism, and can also be achieved by invoking a superselection rule, prohibiting superpositions of states to be written, which is the Sudarshan approach. Sudarshan stipulates that only the positive square root $\sqrt{\rho(\varphi)}$ of the probability density is physically relevant \cite{sud}. One can understand this in terms of the structure of Hilbert spaces in the following way. Since the absolute phase of a state is not observable either in quantum mechanics or in classical optics, all states which differ only by a phase are identified to get `projective rays', resulting in the space $CP = H/U(1)$ of rays. Examples are the Bloch sphere in quantum mechanics and the Poincare sphere in classical polarization optics. However, relative phases are observable in both quantum mechanics and classical optics, but not in classical mechanics. Hence, a further projection or identification is required in classical mechanics, namely, all states that differ by relative phases are to be identified, resulting in the quotient space $CP^* = CP/U(1)$ which implements the `superselection rule'. 

Since a classical electromagnetic field is formally an infinite collection of classical harmonic oscillators and shows interference effects, it is possible to develop a KvN type phase space theory of such fields by using a projective Hilbert space $CP$ so that no superselection rule operates \cite{raja}. Such a field obviously has coherence properties like those of quantum mechanics (including entanglement and Bell violation) except those specifically associated with quantization. This adds a significant new twist to the nature of the wave function---it is not something exclusive to quantum mechanics. 

\section{Acknowledgement}
I thank the National Academy of Sciences, India for the grant of a Senior Scientist Platinum Jubilee Fellowship that enabled this work to be undertaken.

\end{document}